\documentstyle[art12,bezier]{article}

\begin{document}

\renewcommand{\thefootnote}{\fnsymbol{footnote}}

\thispagestyle{empty}

\hfill \parbox{45mm}{
{\sc CTP\ \#2217} \par
June 1993}

\vspace*{15mm}

\begin{center}
{\LARGE A Formula for the Static Potential Energy \par
        in Quantum Gravity.}

\vspace{22mm}

{\large Giovanni Modanese}%
\footnote{On leave from University of Pisa, Pisa, Italy. This work is
supported in part by funds provided by the U.S. Department of Energy
(D.O.E.) under contract \#DE-AC02-76ER03069.}

\medskip

{\em Center for Theoretical Physics \par
Laboratory for Nuclear Science \par
Department of Physics \par
Massachusetts Institute of Technology \par
Cambridge, Massachusetts, 02139, U.S.A.}



\bigskip

\end{center}

\vspace*{20mm}

\renewcommand{\thefootnote}{\arabic{footnote}}
\setcounter{footnote} 0
\begin{abstract}
We give a general expression for the static potential energy of the
gravitational interaction of two massive particles, in terms of an invariant
vacuum expectation value of the quantized gravitational field. This formula
holds for functional integral formulations of euclidean quantum gravity,
regularized to avoid conformal instability. It could be regarded as the
analogue of the Wilson loop for gauge theories and allows in principle, through
numerical simulations or other approximation techniques, non perturbative
evaluations of the potential or of the effective coupling constant. The
geometrical meaning of this expression is quite simple, as it represents the
``average proper-time delay'', respect to two neighboring lines, of a very long
geodesic with unit timelike tangent vector.

\bigskip \bigskip

\end{abstract}

\newcommand{\beq}{\begin{equation}}    \newcommand{\la}{\langle}
\newcommand{\eeq}{\end{equation}}      \newcommand{\ra}{\rangle}
\newcommand{\beqa}{\begin{eqnarray}}   \newcommand{\pa}{\partial}
\newcommand{\eeqa}{\end{eqnarray}}     \newcommand{\half}{\frac{1}{2}}

\newcommand{\m}{\medskip}

\section{Introduction.}
\label{introd}

The present paper is concerned with the problem of the energy of the
gravitational field. This energy has been under investigation
since the birth of General Relativity and some issues, like the
determination of the total energy of a field configuration, have been
settled in a rigorous way in the ADM formalism \cite{dewitt} or
through Noether's theorem \cite{jacki2}. Other points, however, like the
possibility of ``localizing'' the gravitational energy, are still
obscure. Before presenting our contribution, which concerns in fact
the particular issue of the static potential energy, we shall briefly
review a few general facts.

\m
In principle, the energy of the gravitational field is physically as
important as the energy of any other field. In fact, one of the basic
principles of relativistic field theories is that any interaction between two
particles is not instantaneous, but it is transmitted by a field which
propagates with finite velocity. Suppose that the two particles exchange an
amount of energy $E$, as a result of their interaction. If the first particle
loses the energy $E$ at the time $t$, and the second particle receives that
energy at the time $t+\Delta t$, it is usually assumed that in the interval
$(t,\ t+\Delta t)$ the energy is ``stored'' in the field that carries the
interaction.

In practice, however, the gravitational energy turns out to be
much more ``elusive'' than other forms of energy, in the sense that
it seems not possible to ``localize'' it.

\m
For example, the electromagnetic field is known to possess
the local energy density
\beq
  \tau^{00}(x) = \half \left[ {\bf E}^2(x) + {\bf B}^2(x)
  \right] ,
\eeq
which is a component of the energy-impulse tensor
\beq
  \tau^{\mu \nu} = F^{\mu \alpha} F^\nu_{\ \alpha}
  - \frac{1}{4} \eta^{\mu \nu} F^{\alpha \beta} F_{\alpha \beta} .
\label{kkd}
\eeq
(See \S\ \ref{newcon} for our convention about the signature
of the metric and others.)
The tensor $\tau^{\mu \nu}$ is locally conserved, that is
\beq
  \pa_\mu \tau^{\mu \nu} = 0 .
\label{nhw}
\eeq
This assures that the spatial integral of $\tau^{00}$ is conserved,
provided {\bf E} and {\bf B} vanish sufficiently fast at infinity.

\m
Eq.\ (\ref{kkd}) has suggested a generalization
of the tensor $\tau^{\mu \nu}$ to the gravitational
case, called the ``Bel superenergy tensor'' \cite{zakhar}. It has the form
\beq
  {\cal T}^{\mu \nu \rho \sigma} =
  R^{\mu \cdot \nu}_{\cdot \alpha \cdot \beta}
  R^{\rho \alpha \sigma \beta} +
  {R^*}^{\mu \cdot \nu}_{\cdot \alpha \cdot \beta}
  {R^*}^{\rho \alpha \sigma \beta} ,
\label{xkg}
\eeq
where ${R^*}^{\rho \alpha \sigma \beta}$ is a suitably defined
dual of the Riemann tensor.

The tensor ${\cal T}^{\mu \nu \rho \sigma}$ has a positivity property, is
covariantly conserved and allows to define a criterium for the presence of a
gravitational wave in a point of empty spacetime. We remind that, due to the
equivalence principle, the curvature tensor is the true physical field in
General Relativity, since we cannot set it equal to zero at a given point by a
transformation of the coordinates. So it could be fairly expected that a local
energy density contains this tensor. However, the tensor ${\cal T}^{\mu \nu
\rho \sigma}$ defined in (\ref{xkg}) is quadratic in the curvature, and this
prevents us from relating it to the Hamiltonian of General Relativity.

\m
Another problem for a local energy density
resides in the fact that the ordinary conservation equation
of a tensor (like (eq.\ \ref{nhw})) is not generally covariant, and the
introduction of the covariant derivative spoils in fact the conservation. All
we can obtain in General Relativity is thus a conserved energy-impulse
``pseudotensor'' \cite{weinbe}. It leads to an energy which is the integral
of some derivatives of the metric on a surface at spatial infinity, namely
\beq
  P^0 = - \frac{1}{16 \pi G} \int
  \left( \frac{\pa h_{jj}(x)}{\pa x^i} - \frac{\pa h_{ij}(x)}{\pa x^j}
  \right) n^i r^2 d\Omega ,
\label{sle}
\eeq
where it is assumed that the metric is quasi-minkowskian at
infinity and is decomposed as $g_{\mu \nu}(x)=\eta_{\mu \nu}+h_{\mu \nu}(x)$.

\m
The most natural way to obtain the pseudotensor is to apply Noether's theorem
to the case of coordinates transformations, whose generators are the Lie
derivatives \cite{jacki1}. The nice formalism of ``improvement'' of the
Noether currents also works in this case \cite{jacki2}; we symmetrize the
tensor obtained from translational invariance and we finally obtain for the
energy density a quantity  which amounts to a spatial divergence and reduces to
(\ref{sle}) upon spatial integration. This result is typical of the application
of Noether's theorem to gauge symmetries, which usually produce ``trivially
conserved currents'' of the form
\beq
  J_\mu = \pa_\nu \, f^{\mu \nu} ,
\eeq
where $f^{\mu \nu}$ is an antisymmetric tensor.

For known asympotically flat metrics like the Schwartzschild metric,
(\ref{sle}) reproduces the total energy obtained from the ADM hamiltonian
\cite{dewitt,faddee}. (See also ref.\ \cite{haywar} for a comparison of
different ``quasilocal energies'' and references.)

\m
The purpose of this paper is to relate the potential energy of the
gravitational interaction of two particles to an invariant quantum-field
average, which might play in gravity the same role the Wilson loop plays in the
usual gauge theories.

In other words, this energy turns out to be related in a general fashion to the
vacuum average of a simple and well-defined invariant functional of the field.
While evaluation of this average for a weak field on a flat background yields
the usual Newton potential energy, non-perturbative evaluations of the same
average are likely to give rise to modifications in the coupling constant or in
the dependence of the energy on the distance between the particles. We shall
prove, however, that the energy is always negative.

Like in the case of the Wilson loop, our formula can be implemented quite
naturally on a lattice version of the theory, in order to allow numerical
computations.

\m
The outline of the paper is the following.
In \S \ \ref{newcon}, mainly in order to fix our conventions, we compute the
Newton potential starting from the graviton propagator.
In \S \ \ref{intera} we recall the connection between the ADM mass formula
and the static gravitational potential. Through the ADM formula it is
possible, in principle, to find the relativistic corrections to the
Newton potential. In \S\ \ref{quafor} we give a formula which, treating two
masses as external sources for a quantized gravitational field on a flat
background, allows to write the potential energy of their interaction. This is
done using a known technique of euclidean quantum field theory \cite{symanz}.
Such a formula would allow to compute the quantum corrections to the Newton
potential. In \S\ \ref{parlin} the definition of the external source is
generalized, avoiding use of the background metric and introducing the idea of
nearby parallel lines in curved space. Finally, in \S\ \ref{genfor} we
generalize the formula for the potential energy to the case of
``strong'' gravity and in \S\ \ref{appcon} we
suggest some possible applications.


\section{Newton potential. Conventions.}
\label{newcon}

In this paper we work in (3+1) dimensions and follow the conventions of
Weinberg \cite{weinbe}.
Our metric has signature $(-1,\, 1,\, 1,\, 1)$. The Einstein action is given by
\beq
  S_{\rm Einst.} = - \frac{1}{16 \pi G}
  \int d^4x \, \sqrt{g(x)} \, R(x)
\label{skv}
\eeq
and the action of a material particle of mass $m$ is
\beq
  S_{\rm Mat.} = - m \int dp \, \sqrt{-g_{\mu \nu}[x(p)]
  \dot{x}^\mu(p) \dot{x}^\nu(p)} ,
\eeq
where $x^\mu(p)$ is the trajectory of the particle and $p$ is any
parameter. The dots will always denote differentiation with respect to
the parameter. Finally, the Einstein equations have the form
\beq
  R_{\mu \nu} - \half g_{\mu \nu} R = - 8 \pi G T_{\mu \nu} .
\eeq

Let us decompose the metric in the traditional way
\beq
  g_{\mu \nu}(x) = \eta_{\mu \nu} + h_{\mu \nu}(x)
\eeq
and denote the linearized Einstein equations in the harmonic gauge as
\beq
  K^{\rho \sigma}_{\mu \nu} \, h_{\rho \sigma} = T_{\mu \nu} .
\eeq
The inverse of the kinetic operator $K$ is the well-known
Feynman-De Witt propagator
\beq
  K^{-1}_{\mu \nu \rho \sigma}(x-y) =
  \la h_{\mu \nu}(x) h_{\rho \sigma}(y) \ra =
  - \frac{2G}{\pi} \,
  \frac{\eta_{\mu \rho} \eta_{\nu \sigma} +
  \eta_{\mu \sigma} \eta_{\nu \rho} -
  \eta_{\mu \nu} \eta_{\rho \sigma}}{(x-y)^2 - i\epsilon} .
\eeq

\m
Let us now compute the Newton potential starting from the preceding equations.
The field produced by a generic four-momentum source $T_{\rho \sigma}$ in the
linearized approximation is given by
\beq
  h_{\mu \nu}(x) = \int d^4y \, [K^{-1}]_{\mu \nu}^{\rho \sigma}(x-y)
  \, T_{\rho \sigma}(y) ;
\eeq
when the source is a particle of mass $m$ at rest in the origin, the
only non vanishing component of $T_{\rho \sigma}$ is
\beq
  T_{00}(y) = m \, \delta^3({\bf y}) ,
\eeq
so we have
\beqa
  h_{00}(x^0,{\bf x}) & = &
  \int d^4y \, [K^{-1}]^{00}_{00}(x-y) \, m \, \delta^3({\bf y}) \nonumber \\
  & = & m \int_{-\infty}^{+\infty} dy^0 \int d^3y \,
  \frac{\frac{-2G}{\pi} \, \delta^3({\bf y})}
  {({\bf x}-{\bf y})^2-(x^0-y^0)^2-i\epsilon}
  \nonumber \\
  & = & m \int_{-\infty}^{+\infty} dy^0 \,
  \frac{\frac{-2G}{\pi}}{{\bf x}^2-(x^0-y^0)^2-i\epsilon}
  = \frac{2mG}{|{\bf x}|} .
\label{fre}
\eeqa
This is the correct result, since in the newtonian approximation
we have
\beq
  g_{00} = -1 - 2V ,
\eeq
where $V$ is the Newton potential. We shall encounter the integral
appearing in (\ref{fre}) also in \S \ \ref{quafor}.


\section{Potential Energy Versus ADM Energy.}
\label{intera}

In classical General Relativity the total energy (mechanical + gravitational)
of a physical system is given by the ADM mass formula (\ref{sle}), which
has the remarkable property of involving only the gravitational field on
a surface at spatial infinity. We shall briefly recall here the connection
between the ADM energy and the static gravitational potential.

A generic static metric $g_{\mu \nu}$ can be written at spatial
infinity in the form
\begin{eqnarray}
  g_{00} & \simeq & -1 + \frac{2M_1 G}{|{\bf x}|} +
  O\left(\frac{1}{|{\bf x}|^2}\right) ; \label{due} \\
  g_{0i} & \simeq & O\left(\frac{1}{|{\bf x}|^2}\right) ; \label{duf} \\
  g_{ij} & \simeq & \delta_{ij} + \frac{2M_2 G}{|{\bf x}|^3} x_i x_j
  + O\left(\frac{1}{|{\bf x}|^2}\right) . \label{dug}
\end{eqnarray}

Performing the integral (\ref{sle}) with $h_{ij}$ given by (\ref{dug})
one sees that $M_2$ is the ADM energy (``total mass''). On the other hand,
$M_1$ is the mass observed by measuring the newtonian force at infinity.
Substituting (\ref{due}) - (\ref{dug}) into Einstein's equations
$R_{\mu \nu}=0$, it is easy to see that $M_1=M_2$. This is a quite
natural result \cite{dewitt}. In other words it means that, according to
special relativity conceptions, the source of the newtonian
field is not only the mass, but also the energy density.
For instance, in the gravitational collapse of a star a part of the
gravitational energy is converted into kinetic energy and eventually
this energy is employed to produce heavier elements from hydrogen or
helium. If we disregard the radiation emitted into space, the newtonian
field far away from the star remains unchanged during the whole process
and the same holds for the ADM mass, which is a conserved quantity.

The gravitational potential energy can be found, by definition, assuming
a static distribution of matter and computing the metric it generates at
infinity. This has been done by Murchada and York \cite{muryor} for a
spherical matter distribution of uniform density, using conformal
transformations and a special formulation of the initial-value equations
of General Relativity. For a sphere of (small) density $\rho$ and unit
radius they found the right newtonian gravitational binding energy,
namely the ADM mass is given in this case by (reintroducing the radius
$R$ and the velocity of light $c$)
\begin{equation}
  M_{\rm TOT} = \frac{4}{3} \pi R^3 \rho - \frac{1}{c^2} \frac{16}{15}
  \pi^2 \rho^2 G R^5 + O(\rho^3) .
\label{xsw}
\end{equation}
Remembering that $M_{\rm TOT}$ also represents the effective source of the
newtonian field, we see that the second term in (\ref{xsw}) gives
rise to a deviation from the famous law which states the independence
of the potential on the radius of the source. Nevertheless, this
effect is usually unobservable, due to the very small factor $c^{-2}$.

It is also possible to find the following corrections to (\ref{xsw}),
proportional to $\rho^3$, $\rho^4$, ... They denote the existence of
general-relativistic corrections to the potential energy $m_1 m_2 G/r$.
For instance, the term proportional to $\rho^3$ would contribute to
$M_{\rm TOT}$ a term of the form
\begin{equation}
  \Delta M \propto \frac{1}{c^4} \rho^3 G^2 R^7 + O(\rho^4).
\label{kgf}
\end{equation}

In the case of a source constituted by two pointlike bodies of masses
$m_1$ and $m_2$, kept at rest at a fixed distance $r$, the method of
solution mentioned above is not applicable. (An approximate solution
is still possible \cite{sdeser}.) From the preceding discussion we may
infer that the ADM mass is given in this case by
\begin{equation}
  M_{\rm TOT} = m_1 + m_2 - \frac{1}{c^2} \frac{G m_1 m_2}{r} +
  o\left(\frac{1}{c^2}\right).
\end{equation}
We are not able, however, to deduce the relativistic corrections to the
two-body potential from (\ref{kgf}), because the corresponding potential
does not admit a continuum limit. For instance, if we try to integrate
a potential of the form $G^2 m_1^{3/2} m_2^{3/2}/r^2$ to obtain the
term proportional to $\rho^3$, we find that the binding energy of the
sphere depends on the way it has actually been put together.


\section{Quantum formula for the potential energy on
a flat background.}
\label{quafor}

The same result we found in the preceding Section using the classical equations
of motion can be obtained in a completely different way. It is known
that the ground state energy of a system described by an action
$S_0[\phi]=\int d^4x \, L(\phi(x))$ in the presence of external sources $J(x)$
can in euclidean quantum field theory be expressed as
\beq
  {\cal E} = \lim_{T \to \infty} - \frac{1}{T} \log
  \frac{\int d[\phi] \exp \left\{ -\int d^4x \, L(\phi(x)) +
  \int d^4x \, \phi(x) J(x) \right\}}{\int d[\phi]
  \exp \left\{ -\int d^4x \, L(\phi(x)) \right\} } ,
\eeq
where, outside the interval $(-\half T,\ \half T)$, the source has been
switched off.

\m
This formula has been proved exactly in perturbation theory \cite{symanz} for
the case of a linear local coupling between the field and the external source,
but it can be generalized if we assume that in any case the vacuum-to-vacuum
transition amplitude is given by
\beq
  \la 0^+ | 0^- \ra_J =
  \frac{\int d[\phi] \exp \left\{ -S_0[\phi] + S_{\rm Inter.}[\phi,J] \right\}}
  {\int d[\phi] \exp \left\{-S_0[\phi] \right\} } .
\eeq
In fact, inserting a complete set of energy eigenstates we can write
\beq
  \la 0^+ | 0^- \ra_J = \la 0 | e^{-HT} | 0 \ra =
  \sum_n \la 0 | e^{-HT} | n \ra \la n | 0 \ra =
  \sum_n |\la 0 | n \ra|^2 \, e^{-E_n T} .
\eeq
The smallest energy eigenvalue $E_n$ corresponds to the ground state, and in
the limit $T \to \infty$ it dominates the sum. So taking the logarithm and
dividing by $(-T)$ we obtain that energy. This is a well-known technique in QCD
(see for instance \cite{bander}).

\m
In the case of a weak gravitational field quantized on a flat background, we
may consider the source constituted by two masses $m_1,\ m_2$, placed at rest
near the origin at a distance $L$ each from the other (see eq.\ (\ref{kty})).
The action of this system is
\beqa
  S & = & S_{\rm Einst.} + S_{\rm Mat.,1} + S_{\rm Mat.,2} = \nonumber \\
  & = & - \frac{1}{16\pi G} \int d^4x \sqrt{g(x)} \, R(x)
  - m_1 \int_{-\frac{T}{2}}^{\frac{T}{2}} dt_1 \,
  \sqrt{-g_{\mu \nu}[x(t_1)] \dot{x}^\mu(t_1) \dot{x}^\nu(t_1) }
  \nonumber \\
  & & - m_2 \int_{-\frac{T}{2}}^{\frac{T}{2}} dt_2 \,
  \sqrt{-g_{\mu \nu}[y(t_2)] \dot{y}^\mu(t_2) \dot{y}^\nu(t_2) } ,
\eeqa
where the trajectories $x^\mu(t_1)$ and $y^\mu(t_2)$ of the
particles with respect to the background are simply given by
\beq
  x^\mu(t_1) = \left( t_1,\, -\frac{L}{2},\, 0,\, 0 \right); \qquad
  y^\mu(t_2) = \left( t_2,\, \frac{L}{2},\, 0,\, 0 \right) .
\label{kty}
\eeq

So we have, denoting by $\hat{S}$ the euclidean action,
\beqa
  & & {\cal E} = \lim_{T \to \infty} - \frac{1}{T} \times \nonumber \\
  & & \ \ \log \frac{\int d[h] \, \exp \left\{ - \hat{S}_{\rm Einst.} -
  m_1 \int dt_1 \,
  \sqrt{1 - h_{00}[x(t_1)] } -
  m_2 \int dt_2 \,
  \sqrt{1 - h_{00}[y(t_2)]} \right\} }
  {\int d[h] \, \exp \left\{ - \hat{S}_{\rm Einst.} \right\} } .
  \nonumber \\ & &
\label{skt}
\eeqa

Since it is known that the euclidean Einstein action is not bounded from below,
due to the existence of ``conformal singularities'' (see \cite{greens}), we
actually mean by $\hat{S}_{\rm Einst.}$ a regularized version of (\ref{skv}).
Such a regularization can be achieved adding $R^2$-terms to the original
Einstein lagrangian \cite{hamber} or through a more recent technique, called
``stochastic regularization'' \cite{greens}. We shall return to this point in
\S\ \ref{appcon}.

\m
Returning to (\ref{skt}), by standard perturbation techniques it is
straightforward to see that for weak fields and to lowest order in $G$ it
reduces to
\beqa
  {\cal E} & = & m_1+m_2+ \lim_{T \to \infty} - \frac{1}{T} \log
  \left\{ 1 + \frac{m_1 m_2}{4} \int_{-\frac{T}{2}}^{\frac{T}{2}} dt_1
  \int_{-\frac{T}{2}}^{\frac{T}{2}} dt_2 \, \left<
  h_{00}[x(t_1)]
  h_{00}[y(t_2)] \right> \right\}
  \nonumber \\
  & \simeq & m_1+m_2+ \lim_{T \to \infty} - \frac{1}{T} \frac{m_1 m_2}{4}
  \int_{-\frac{T}{2}}^{\frac{T}{2}} dt_1
  \int_{-\frac{T}{2}}^{\frac{T}{2}} dt_2 \,
  K^{-1}_{0000}(\tau_1-\tau_2,\, L,\, 0,\, 0) \nonumber \\
  & = & m_1+m_2 -  \frac{m_1 m_2 G}{L} .
\eeqa
The next term in the perturbative series is the first quantum correction,
proportional to $\hbar G^2$. We shall compute it exactly in a forthcoming
paper.

\m
Eq.\ (\ref{skt}), like the corresponding ones in QED or QCD, has the physically
appealing feature of showing how the force between the sources ultimately
arises from the exchange of massless bosons. However, let us make a closer
comparison with electrodynamics. In that case the analogue of the functional
integral which appears in the logarithm of (\ref{skt}) has the form
\cite{fischl}
\beq
  \left< \exp \left\{ g \int_{-\frac{T}{2}}^{\frac{T}{2}} dt_1
  A_0[x(t_1)] - g \int_{-\frac{T}{2}}^{\frac{T}{2}} dt_2
  A_0[y(t_2)] \right\} \right> .
\label{vge}
\eeq
(The two charges have been chosen to be opposite: $q_1=g$, $q_2=-g$.) Reversing
the direction of integration in the second integral and closing the contour at
infinity, one is able to show that the quantity (\ref{vge}) coincides with the
Wilson loop of a single charge $g$, thus giving a gauge invariant expression
for the potential energy.

In gravity this is not possible: we may imagine that an expression like
(\ref{vge}) could be obtained in the first-order formalism (with $A_0$ replaced
by the tetrad $e^0_0$), but the masses necessarily have the same sign.

\m
Luckily, our formula for the energy is invariant as it stands, as we shall
see better in the next Sections, where we generalize it to the case where no
background metric is fixed.


\section{Parallel lines in curved space.}
\label{parlin}

In order to generalize eq.\ (\ref{skt}) beyond the case of weak fluctuations of
the gravitational field around a fixed flat background, we need a definition of
the source that does not depend on such a background.

\m
We assume that a functional integral for regularized euclidean
gravity exists, denoted by
\beq
  z = \int d[g] \, \exp \left\{ - \hat{S}_{\rm Einst.}[g] \right\} ,
\eeq
and we require that all the field configurations in this functional integral
are asymptotically flat.

\m
Let us suppose that a field configuration is given. We consider a {\it geodesic
line of length $T$, which starts at an arbitrary point in the ``past''
asymptotically flat region with unit timelike velocity}.

To fix the ideas, this curve could be written in its first part as
\beq
  \xi^\mu(\tau) = \left( -\half T + \tau, \, 0,\, 0,\, 0 \right) ;
  \qquad 0 \leq \tau \leq \bar{\tau} ,
\eeq
where $\tau$ is the proper time measured along the curve and we have chosen the
spatial coordinates of the starting point to be equal to $(0,\, 0,\, 0)$ (this
is an irrelevant arbitrariness, since at the end we shall integrate over all
the configurations of the field). As usual, $T$ denotes a very long time
interval. After a time $\simeq \bar{\tau}$ the curve enters the region of
spacetime where the gravitational field is non vanishing. It continues as a
geodesic, which means that $\xi^\mu(\tau)$ satisfies the equation
\beq
  \Gamma^\rho_{\mu \nu}[\xi(\tau)] \dot{\xi}^\mu(\tau) \dot{\xi}^\nu(\tau)
  + \ddot{\xi}^\rho(\tau) = 0,
\eeq
where $\Gamma^\rho_{\mu \nu}$ is the Christoffel symbol of the metric. The
curve terminates at $\tau=\half T$, again in the flat region.

\m
Let us then take in the initial point $\xi^\mu(0)$ a unit vector $q^\mu(0)$,
orthogonal to $\dot{\xi}^\mu(0)$ (for instance, in our preceding example,
$q^\mu(0)=(0,\, 1,\, 0,\, 0)$), and define a vector $q^\mu(\tau)$ along the
curve $\xi^\mu(\tau)$ by parallel transport of $q^\mu(0)$. We remind that
$\dot{\xi}^\mu(\tau)$, being the tangent vector of a geodesic, is parallel
transported along the geodesic itself, and that the parallel transport
preserves the norms and the scalar products. Then the following relations
hold along the curve
\beqa
  \dot{\xi}^\mu(\tau) \dot{\xi}^\nu(\tau) g_{\mu \nu}[\xi(\tau)] & = & -1 ;\\
  q^\mu(\tau) q^\nu(\tau) g_{\mu \nu}[\xi(\tau)] & = & 1 ;\\
  \dot{\xi}^\mu(\tau) q^\nu(\tau) g_{\mu \nu}[\xi(\tau)] & = & 0 .
\eeqa

Next we consider two masses $m_1, \ m_2$, and a length $L$ which we may regard
as infinitesimal, compared to the scale $T$. We assume that the two masses
follow the trajectories $x^\mu(\tau)$ and $y^\mu(\tau)$, respectively, given by
\beqa
  x^\mu(\tau) & = & \xi^\mu(\tau) - L_1 q^\mu(\tau) ; \label{zsw} \\
  y^\mu(\tau) & = & \xi^\mu(\tau) + L_2 q^\mu(\tau) , \label{zss}
\eeqa
where $L_1$ and $L_2$ are two positive lengths such that
\beq
  L_1+L_2 = L \qquad {\rm and} \qquad -m_1 L_1 + m_2 L_2 = 0 .
\label{huy}
\eeq

\m
The physical meaning of the preceding geometrical construction is apparent: it
represents an observer which falls freely in the center of mass of the system
composed by $m_1$ and $m_2$, while holding the two masses at rest at a distance
$L$ each from the other. This is a generalization of the source introduced in
\S\ \ref{quafor} that is naturally dictated by the equivalence principle.

We notice that if the two masses were allowed to fall freely in the field, they
would not keep at a constant distance from each other. In fact, as it is well
known from the so-called geodesic deviation equation, the distance between two
neighboring geodesics varies according to the sign of the curvature in the
region they are traversing.

\m
We can reparameterize the two curves $x^\mu(\tau)$ and $y^\mu(\tau)$
introducing their proper times $\tau_1$ and $\tau_2$, respectively.
The ratio between the proper time $\tau_1$ and the proper time $\tau$
is given by the equation
\beq
  d\tau_1 = \sqrt{-g_{\mu \nu}[x(\tau_1)]
  \dot{x}^\mu(\tau_1) \dot{x}^\nu(\tau_1)} d\tau ,
\eeq
where $\tau_1=\tau_1(\tau)$; using (\ref{zsw}), (\ref{zss}), we have
\beqa
  \left( \frac{d\tau_1(\tau)}{d\tau} \right)^2 & = &
  1 + L_1 \, q^\alpha(\tau) \pa_\alpha g_{\mu \nu}[\xi(\tau)] \,
   \dot{\xi}^\mu(\tau) \dot{\xi}^\nu(\tau) + \nonumber \\
  & & + L_1 \, g_{\mu \nu}[\xi(\tau)] \left\{ \dot{\xi}^\mu(\tau)
  \dot{q}^\nu(\tau) + \dot{\xi}^\nu(\tau) \dot{q}^\mu(\tau)
  \right\} + O(L_1^2) .
\label{bkw}
\eeqa
An analogous relation holds for $\tau_2$. We agree to adjust the function
$\tau_1(\tau)$ in such a way that $\tau_1(0)=0$. Then we shall denote
$\tau_1(-\half T)=-\half T_1'$ and $\tau_1(\half T)=\half T_1''$. For flat
geometries we have $T_1'=T_1''=T$. Analogous relations hold for $\tau_2$.

\m
We notice that eq.\ (\ref{bkw}) takes a much simpler form in the coordinate
system where $\xi^\mu(\tau)$ defines one axis of a normal coordinates system
(if they can be globally defined). In that case, we have
$\dot{\xi}^\mu(\tau)=(1,\, 0,\, 0,\, 0)$, $\dot{q}^\mu(\tau)=0$ and (\ref{bkw})
reduces to
\beq
  \left( \frac{d\tau_1}{d\tau} \right)^2 = 1 + L_1 \pa_1
  g_{00}(\tau,\, 0,\, 0,\, 0) + O(L_1^2) ,
\eeq
which obviously reminds us of the equations of \S\ \ref{quafor}. However, we
are interested only in coordinates-independent quantities in the following.


\section{General formula for the potential energy.}
\label{genfor}

According to the discussion of the preceding Section, in the absence
of a background eq.\ (\ref{skt}) must be rewritten as
\beqa
  {\cal E} & = & \lim_{T \to \infty} - \frac{1}{T} \log
  \left< \exp \left\{ - m_1 \int_{-\half T_1'}^{\half T_1''}
  d\tau_1 \, \sqrt{-g_{\mu \nu}[x(\tau_1)]
  \dot{x}^\mu(\tau_1) \dot{x}^\nu(\tau_1) } \right. \right. \nonumber \\
  & & \qquad \qquad \qquad \ \
  \left. \left. - m_2 \int_{-\half T_2'}^{\half T_2''}
  d\tau_2 \, \sqrt{-g_{\mu \nu}[y(\tau_2)]
  \dot{y}^\mu(\tau_2) \dot{y}^\nu(\tau_2) } \right\}
  \right>_{\hat{S}_{\rm Einst.}} \\
  & & \nonumber \\
  & = & \lim_{T \to \infty} - \frac{1}{T} \log
  \left< \exp \left\{ - \half m_1 \left( T_1' + T_1'' \right)
  - \half m_2 \left( T_2' + T_2'' \right) \right\}
  \right>_{\hat{S}_{\rm Einst.}} ,
\eeqa
where, for brevity, we have denoted by brackets the functional
average weighted by the exponential of the (regularized) Einstein action.

\m
Now we exploit the property, characteristic of timelike geodesics in a
Lorentzian manifold, of having maximal length with respect to neighboring
lines. This means that
\beqa
  \half m_1 \left( T_1' + T_1'' \right) & = &
  T \left\{ 1 - \delta_1 \left( L_1, [g] \right) \right\} ; \\
  \half m_2 \left( T_2' + T_2'' \right) & = &
  T \left\{ 1 - \delta_2 \left( L_2, [g] \right) \right\} ,
\eeqa
where $\delta_{1,2}$ are small positive adimensional functionals of the
geometry $g$, which also depend on $L_{1,2}$ and thus, -- through eq.\
(\ref{huy}) -- on $L$ and $m_1$, $m_2$. So we have
\beqa
  {\cal E} & = & \lim_{T \to \infty} - \frac{1}{T} \log
  \left< e^{ - T (m_1 + m_2)}
  \exp \left\{ T \left( m_1 \delta_1 \left( L_1, [g] \right) +
  m_2 \delta_2 \left( L_2, [g] \right)
  \right) \right\} \right>_{\hat{S}_{\rm Einst.}} \nonumber \\
  & \simeq & m_1 + m_2 - \left< m_1 \delta_1 \left( L_1, [g] \right) +
  m_2 \delta_2 \left( L_2, [g] \right) \right>_{\hat{S}_{\rm Einst.}} .
\label{epi}
\eeqa
The quantity in the bracket represents the energy of the gravitational
interaction. More exactly, it represents,
apart from $m_1$, $m_2$, the ground state energy of the system
constituted by the source described in \S\ \ref{parlin}, coupled to a quantum
gravitational field. By construction, ${\cal E}$ is invariant with respect to
coordinate transformations.

\m
We see from eq.\ (\ref{epi}) that the interaction energy is always negative.
Apart from this, (\ref{epi}) alone does not give us any precise indication on
the dependence of this energy on the masses, on $L$ and on $G$ (which is
contained in the action). This dependence is a nontrivial result of the
dynamics, as we have seen already in the simple perturbative example of \S\
\ref{quafor}. We can apply to the functional integral appearing in (\ref{epi})
other approximation techniques, like the semiclassical approximation, or
discretize it and use numerical methods.

In any case, the geometrical meaning of (\ref{epi}) is quite simple. In
practice, this formula implies the following: (1) trace a geodesic with unit
timelike tangent vector through one (asympotically flat) field configuration;
(2) measure the total ``delay'' of this geodesic with respect to two
neighboring lines; (3) average on many configurations.


\section{Possible applications and concluding remarks.}
\label{appcon}

The functional integral representation for the static gravitational potential
given in (\ref{epi}) can serve as a basis for various approximations
(weak-field, semiclassical) and for numerical simulations. The latter require a
discretized version of the theory in which the geodesic lengths or the metric
represent the fundamental variables. The most suitable technique under this
respect is probably the ``quantum Regge calculus'' of Hamber \cite{hamber};
another candidate, which might have the advantage of simpler algorithms, is the
``stochastic stabilized gravity'' of Greensite \cite{greens}. As we pointed out
in the last Section, the geometrical meaning of our formula is quite simple.
This should allow us to write a dedicated algorithm for its evaluation, in
order
to improve the efficiency of the method.

\m
The most simple quantity to be ``measured'' in this way in a lattice version of
gravity is the effective coupling constant $G$. In order to compute it through
${\cal E}$, we just need to set $m_1=m_2$, $L_1=L_2$ and assign some fixed
values to them, so that the algorithm can be even more simplified.

There have been some suggestions (see for ex.\ \cite{polyak}) that the
effective coupling constant in gravity is scale dependent and decreases at
small distances. This ``antiscreening'' of the gravitational interaction seems
to be quite natural, since the longer is the cloud of virtual particles, the
stronger is the gravitational force. In a forthcoming paper we shall
employ our formula to compute the correction of order $\hbar$ to the
potential energy, and possibly to exibit such an effect.

\m
In conclusion, we remind that this paper is part of a program which aims to
study physical observables in four-dimensional quantum gravity. Other
observables considered were the vacuum correlations at goedesic distance
\cite{radcor} and the loops of the Christoffel connection \cite{georou}.


\section{Acknowledgments.}

It is a pleasure to thank prof.\ Roman Jackiw for the kind hospitality at
M.I.T.\ and for charming discussions about the problem of energy in General
Relativity, as well as for helpful criticism and suggestions concerning this
work. I also am grateful to prof.\ S.\ Deser for invaluable clarifications
about the ADM energy. The author is supported by a fellowship of the Foundation
``A. Della Riccia'' of Florence, Italy.


\end{document}